\documentclass[journal=jctcce, manuscript=article, layout=twocolumn, email=false]{achemso}
\setkeys{acs}{articletitle=true}
\usepackage{amsfonts,amssymb,amsmath}
\usepackage[english]{babel}
\usepackage[utf8x]{inputenc}
\usepackage[T1]{fontenc}

\usepackage{threeparttable}
\usepackage{lmodern}
\usepackage{dcolumn}
\usepackage{graphicx}
\usepackage[version=3]{mhchem}
\usepackage{paralist}
\usepackage{lineno,xcolor}
\usepackage{cancel}
\usepackage{siunitx}
\usepackage{float}
\usepackage[colorlinks,
  allcolors=blue,
  citecolor=blue,
  urlcolor=blue,
  bookmarks]{hyperref}

\usepackage{booktabs}
\usepackage{caption,setspace}
\usepackage{tabularx}
\usepackage{multirow}
\usepackage{bm}
\usepackage{placeins}

\DeclareMathOperator{\Tr}{Tr}

\DeclareMathOperator*{\argmin}{arg\,min}
\newcommand{\Matrix}[1]{\ensuremath{\bm{#1}}}

\newcommand{\D}{\ensuremath{\mathrm{d}}}
\newcommand{\imag}{\ensuremath{\mathrm{i}}}

\newcommand{\ra}{\mathbf{r}}
\newcommand{\rb}{\ensuremath{\mathbf{r'}}}

\newcommand{\rab}{|\rb-\ra|}

\newcommand{\veeab}{\upsilon_\text{ee}\left(\rab\right)}

\renewcommand{\vee}{\upsilon_\text{ee}}

\newcommand{\RNum}[1]{\ensuremath{\uppercase\expandafter{\romannumeral #1\relax}}}

\title{Random Phase Approximation Applied to Many-Body Noncovalent Systems}
\author{Marcin Modrzejewski}
\email{m.m.modrzejewski@gmail.com}
\affiliation{Department of Chemical Physics and Optics, Faculty of Mathematics and Physics, Charles University, Ke Karlovu 3, CZ-12116
Prague 2, Czech Republic}
\alsoaffiliation{Faculty of Chemistry, University of Warsaw, Pasteura 1, 02-093 Warsaw, Poland}
\author{Sirous Yourdkhani}
\affiliation{Department of Chemical Physics and Optics, Faculty of Mathematics and Physics, Charles University, Ke Karlovu 3, CZ-12116
Prague 2, Czech Republic}
\author{Jiří Klimeš}
\affiliation{Department of Chemical Physics and Optics, Faculty of Mathematics and Physics, Charles University, Ke Karlovu 3, CZ-12116
Prague 2, Czech Republic}

\begin{document}
\begin{abstract}
The random phase approximation (RPA) has received a considerable
interest in the field of modeling systems where noncovalent interactions
are important.
Its advantages over widely used density functional theory (DFT)
approximations are the exact treatment of exchange and the description
of long-range correlation.
In this work we address two open questions related to RPA.
First, how accurately RPA describes nonadditive interactions encountered
in many-body expansion of a binding energy.
We consider three-body nonadditive energies in molecular and atomic clusters.
Second, how does the accuracy of RPA depend on input provided by different DFT models,
without resorting to selfconsistent RPA procedure which is currently impractical
for calculations employing periodic boundary conditions.
We find that RPA based on the SCAN0 and PBE0 models, i.e., hybrid DFT,
achieves an overall accuracy between CCSD and MP3 on a dataset of
molecular trimers of \v{R}ez\'{a}\v{c} et al.
({\it J. Chem. Theory. Comput.} {\bf 2015}, {\it 11}, 3065)
Finally, many-body expansion for molecular clusters and solids often
leads to a large number of small contributions that need to be calculated
with a high precision.
We therefore present a cubic-scaling (or SCF-like) implementation of RPA
in atomic basis set, which is designed for calculations with a high numerical
precision.

\end{abstract}

\maketitle

\section{Introduction}

Molecular solids are materials which are held by noncovalent interactions.
Such molecular solids have often different phases or polymorphs that differ
very little in energy.\cite{price2009computer,nyman2015static}
For example, in more than one half of the polymorph pairs studied by Nyman and Day  
a lattice energy difference was found to be less than 0.5~kcal/mol.\cite{nyman2015static}
There is an ongoing effort to develop theoretical methods that would reliably describe such minute 
differences.\cite{reilly2016report,cervinka2018ab,addicoat2018crystal,hoja2019reliable}
When the binding energy of a solid or a molecular cluster is decomposed using many-body expansion, 
the two-body contribution, corresponding to the binding of dimers, is the largest and as such it
receives most of the attention in method development.
However, nonadditive higher-order contributions, such as three-body and four-body terms can be important as well.
For example, 3-body correlation effects represent some 5 to 7 \% of the correlation contribution 
to the lattice energy in benzene.\cite{kennedy2014resolving,yang2014determination} 
Moreover, they can affect the relative stability of different molecular clusters 
and solid phases.\cite{lotrich1997three,gora2011interaction,gillan2013energy,gillan2016perspective}

Kohn-Sham density functional theory (DFT) has been widely used to understand the binding 
of molecular solids or clusters.
The missing description of long-range correlation (dispersion) has been viewed as the largest 
deficiency of approximate DFT for such systems.
Improving the description of dispersion has therefore been one of most active fields of DFT development
in the past decade.\cite{grimme2016dispersion,vydrov2010nonlocal,lee2010higher,grimme2010consistent,caldeweyher2017extension,ambrosetti2014long}
However, errors originating from the exchange functional are on the same order as the errors originating
from the missing dispersion energy.\cite{gillan2014many,hapka2017nature,jankiewicz2018dispersion}
Approximate exchange functionals alone can lead to both strongly overestimated and underestimated
noncovalent interactions.\cite{lacks1993pair}
For two body systems such issues tend to be masked by adjusting the equilibrium- and short-distance behavior of
the dispersion correction.\cite{gillan2014many}
However, the three-body exchange errors cannot be compensated in a similar way by adjusting the
pairwise additive dispersion corrections.\cite{gillan2014many}
Overall, the conclusion originating from the existing literature is that no existing semilocal functional can 
reliably account for many body effects.\cite{JordanJPCB2014:clathrate,jankiewicz2018dispersion}

Affordable schemes based on perturbation theory could offer higher 
and systematically improvable accuracy for calculations of condensed systems
compared to standard DFT functionals.\cite{jankiewicz2018dispersion,lao2014accurate,huang2015reliable,lao2018atomic,carter2019accurate}
Of such schemes, the random phase approximation to the correlation energy (RPA) is promising
as it is both compatible with the Hartree-Fock (HF) exchange and it contains terms describing
higher-order (nonadditive) correlation effects.\cite{dobson2012dispersion,dobson2012calculation}
RPA has been tested for interaction energies of dimers,\cite{eshuis2011a,bleiziffer2013efficient,ren2013renormalized} 
for adsorption,\cite{torres2017adsorption,alhamdani2017properties,brandenburg2019physisorption} 
or for molecular\cite{lu2009ab,delben2013electron,klimes2016lattice,zen2018fast} 
and atomic solids\cite{harl2008cohesive}
and interfaces.\cite{mittendorfer2011graphene,olsen2013random}
For the cases involving noncovalent interactions, high accuracy has been achieved with addition 
of the singles corrections.\cite{ren2011beyond,klimes2015singles}
However, its accuracy for predicting nonadditive energies is unknown and this is one of our interests in this work.
Moreover, most of the RPA calculations nowadays are performed non-self-consistently, using DFT 
orbitals and energies in the RPA energy expression.
Here we obtain RPA results using different DFT input orbitals.
This allows us to understand if and how the errors in the underlying Kohn-Sham DFT potential 
transfer to the RPA results.

The three- and higher-body energies per single trimer are often small, on the order of microhartree or cal/mol 
or even less for larger distances.
However, in molecular solids such contributions might not be negligible as there can be a large number of them.
Therefore, they need to be evaluated with a high precision so that the resulting value is meaningful.
To this end, we introduce an algorithm for calculating RPA correlation
energies which uses Cholesky decomposition of the Coulomb operator matrix in atomic basis set together
with eigendecomposition of the dielectric matrix.
These steps lead to cubic scaling with the system size and to high and controllable precision. 
The details of the algorithm are presented in Section~2 followed by results on dimers and trimers of noble
gases and results of the 3B-69 test set of Řezáč and coworkers.\cite{rezac2015benchmark}

\section{Theory}
\subsection{Direct RPA}
The adiabatic-connection fluctuation-dissipation formula\cite{dobson2012calculation,dobson2012dispersion}
expresses the exact DFT correlation energy in terms of the density response function $\chi$ integrated over imaginary
frequencies and the adiabatic connection coupling constant $\alpha$:
\begin{multline}
  E_\text{c} = 
-\frac{1}{2 \pi} \int_0^1 \D \alpha \int_0^\infty \D u \int \D^3 \ra \, \D^3 \rb
   \left( \chi^\alpha(\ra, \rb; \imag u) \right. \\
   \left. - \chi^0(\ra, \rb; \imag u) \right) \veeab \,.\label{acfd-formula}
\end{multline}
Here, $v_\text{ee}$ is the electron-electron interaction operator.
The density response function produces a change in the electron density corresponding to a perturbation
of the system's hamiltonian. For the noninteracting system at $\alpha=0$
\begin{equation}
  \delta \rho = \chi^0 \delta \upsilon_s
\end{equation}
where $\delta \upsilon_s$ is the change in the Kohn-Sham effective potential.\cite{hybertsen1987ab}
In this work, the density response function of the Kohn-Sham system is built using orbitals and orbital
energies computed with an approximate exchange-correlation
functional, e.g., the Perdew, Burke, and Ernzerhof (PBE) functional\cite{perdew1996generalized} and SCAN.\cite{sun2015strongly}
An alternative, not explored here, would be to compute self-consistent orbitals.\cite{strubbe2012response,hybertsen1987ab,bleiziffer2013efficient,graf2019low} 
In either case, the formula  for the density response reads
\begin{equation}
  \chi^0(\ra, \rb; \imag u) = 2 \sum_{ia} \frac{ \phi_i(\ra) \phi_a(\ra) \phi_i(\rb) \phi_a(\rb) }
      {\epsilon_i - \epsilon_a + \imag u} + \text{c.c.}        \label{chi-kohnsham}
\end{equation}
The factor of 2 originates from the spin summation over doubly occupied orbitals.
Spin indices should be added for an open shell system. Throughout this work, we assume real orbitals expanded in
an atomic-orbital basis set. Occupied and virtual indices are $i$ and $a$, respectively.
The atomic orbital labels are $p$, $q$, $r$, and $s$.

The interacting density response at $\alpha > 0$ is related to the Kohn-Sham response
by the Dyson-type screening equation\cite{gross1996density}
\begin{equation}
  \chi^\alpha = \chi^0 + \chi^0 \left(\alpha \upsilon_\text{ee} + f_\text{xc}^\alpha \right) \chi^\alpha\,,
\end{equation}
where $f_\text{xc}^\alpha$ is the frequency-dependent exchange-correlation kernel.
The direct random-phase approximation amounts to setting $f_\text{xc}^\alpha=0$,
which leads to
\begin{equation}
  \chi^\alpha = (1 - \alpha \chi^0 \upsilon_\text{ee})^{-1} \chi^0\,. \label{chi-rpa}
\end{equation}
We refer to direct RPA as ``RPA'' for short.
As a consequence of the simple dependence on $\alpha$,
the integration over the coupling constant in the ACFD formula
is done analytically, which yields the final form of the RPA correlation energy
\begin{equation}
  E_\text{c} = \frac{1}{2 \pi} \int_0^\infty 
  \Tr \left( \ln(1 - \chi^0 \upsilon_\text{ee}) + \chi^0 \upsilon_\text{ee} \right) \, \D u \,.\label{rpa-energy}
\end{equation}

To obtain a workable expression for the correlation energy, we define an auxiliary matrix $\Matrix{\Pi}(u)$, which
is a symmetrized matrix representation of the product $-\chi^0 \upsilon_\text{ee}$. 
With that definition, the energy expression reads
\begin{equation}
  E_\text{c} = \frac{1}{2 \pi} \int_0^\infty 
  \Tr \left( \ln(1 + \Matrix{\Pi}(u)) - \Matrix{\Pi}(u)  \right) \, \D u \,.\label{rpa-energy-diel}
\end{equation}
and is in practice evaluated with a numerical quadrature on a frequency grid.

\subsection{Effective basis for the RPA energy}  \label{implementation-description}
We now discuss the technical aspects of our implementation of the ACFD energy formula.
The expression for $\chi^0$, Eq.~\ref{chi-kohnsham}, suggests that it can be represented using
a basis of occupied-virtual orbital pairs $\phi_i(\ra) \phi_a(\ra)$.
In canonical-basis implementations\cite{furche2005fluctuation} the resulting matrix $\chi^0_{ai,bj}$
has a dimension of $N_\text{occ} N_\text{virt}$, implying a significant computational
and storage cost beyond traditional DFT. 
To overcome that issue we use an alternative strategy which combines several algorithms
to reduce the computational cost while not significantly reducing the precision of the result.
First, we represent $\chi^0$ in the basis of atomic orbitals.
This allows us to use the Laplace transform to calculate $\chi^0$ using Green's functions
in imaginary time.
Second, we use Cholesky decomposition of the Coulomb matrix to avoid the use of four-index Coulomb integrals.
Finally, we approximate the auxiliary matrix \Matrix{\Pi} by using only its dominant eigenvectors.
These steps allow for a cubic scaling implementation and a well-defined control of precision. 

Let us first write down the Kohn-Sham density response function as
\begin{multline}
  \chi^0(\ra,\rb;\imag u) = \\
  \sum_{ia} \phi_i(\ra) \phi_a(\ra) \frac{4 (\epsilon_i - \epsilon_a)}{(\epsilon_i - \epsilon_a)^2 + u^2} \phi_i(\rb) \phi_a(\rb) \\
  = -4 \sum_{ia}  \phi_i(\ra) \phi_a(\ra) \frac{d_{ai}}{d_{ai}^2 + u^2} \phi_i(\rb) \phi_a(\rb) \label{real-chi}
\end{multline}
where
\begin{equation}
  d_{ai} = \epsilon_a' - \epsilon_i'
\end{equation}
and the orbital energies $\epsilon_a' = \epsilon_a - \epsilon_\text{F}$ and $\epsilon_i' = \epsilon_i - \epsilon_\text{F}$
are shifted by the Fermi energy $\epsilon_\text{F} = (\epsilon_\text{HOMO} + \epsilon_\text{LUMO}) / 2$.
Following Kaltak et al.,\cite{kaltak2014low} we separate the occupied and virtual indices by
applying the Laplace transform
\begin{align}
  \frac{d_{ai}}{d_{ai}^2 + u^2} &= \int_0^\infty \cos(u t) \exp(-d_{ai} t) \, \D t  \nonumber \\
  &= \int_0^\infty  \cos(u t) \exp(-\epsilon_a' t) \exp(\epsilon_i' t) \, \D t \,.
\end{align}
Now, expanding the molecular orbitals in terms of atomic functions
\begin{align}
  \phi_i(\ra) &= \sum_p C_{pi} \phi_p(\ra) \\
  \phi_a(\ra) &= \sum_p C_{pa} \phi_p(\ra)
\end{align}
yields the density response function expressed completely in terms of AO indices
\begin{equation}
  \chi^0_{pq,rs}(\imag u) = -4 \int_0^\infty \cos(u t) \rho_{pr}^\text{occ}(t) \rho_{qs}^\text{virt}(t) \, \D t
\end{equation}
The matrices $\rho^\text{occ}$ and $\rho^\text{virt}$ are hole and particle noninteracting Green's functions
at imaginary time
\begin{align}
  \rho_{pq}^\text{occ}(t) &= \sum_i C_{pi} \exp(\epsilon_i' t) C_{qi} \\
  \rho_{pq}^\text{virt}(t) &= \sum_a C_{pa} \exp(-\epsilon_a' t) C_{qa}\,.
\end{align}

In principle, the summations over AO pairs $pq$ run
over a range much larger than the original set of occupied-virtual pairs $ia$. However, in a Gaussian
orbital basis set, the Coulomb matrix elements $V_{pq,rs} = \left(pq\middle| \vee \middle|rs\right)$
acquire Gaussian damping factors which decay quickly with the distance between the centers of $pq$ and $rs$.
To take advantage of that, we constrain the computations and storage to the set of significant orbital
shell pairs $\mathcal{S}$. Instead of full summations over $pq$, all sums traverse $\mathcal{O}(N)$
elements of $\mathcal{S}$. The set $\mathcal{S}$ is constructed by discarding
small diagonal Coulomb integrals as long as the trace error of the Coulomb matrix
lies below a predefined error bound
\begin{equation}
  \sum_{pq} V_{pq,pq} - \sum_{pq \in \mathcal{S}} V_{pq,pq} < \tau_\text{screen} \label{significant-set}\,.
\end{equation}
(Note that the bookkeeping is done for whole shell pairs and not for the individual angular functions.)
The prescreening is mostly effective for medium and large systems, which includes noncovalent complexes of
small molecules at separations beyond equilibrium. For example, in an RPA/aug-cc-pVQZ calculation
for methane in a \ce{(H2O)20} cage, the prescreening subroutine
discards $60\%$ of the $6.9 \times 10^6$ orbital pairs at the default level of numerical precision.
We have found that the screening based on the $pq$ and $rs$ AO pairs is
compatible with high precision targets for the total energy (i.e., with the target error
on the order of $10^{-5}$~kcal/mol for the interaction energy).

Multiple techniques exploit the redundancy in the matrix $\Matrix{V}$ to avoid dealing with
four-index Coulomb integrals. Eshuis et al.\cite{eshuis2010fast} and Ren et al.\cite{ren2012resolution}
have applied density fitting with the Coulomb metric in their quartic-scaling RPA implementations.
Wilhelm et al.\cite{wilhelm2016large} and Schurkus and Ochsenfeld\cite{schurkus2016effective} have achieved
cubic and linear scaling, respectively, owing to the sparse matrices appearing in density fitting
based on the overlap metric. While using the overlap metric introduces sparsity, it is orders of
magnitude less precise than fitting with the Coulomb metric or Cholesky decomposition.
The sensitivity of the RPA energy to different choices of the fitting metric
is investigated in Refs.~\citenum{luenser2017vanishing} and~\citenum{ren2012resolution}.

We have found that for our applications, which require extremely precise interaction energies,
the most dependable method of decomposing the Coulomb matrix is the pivoted Cholesky
decomposition\cite{aquilante2011choleski,harbrecht2012low} in which the Coulomb
matrix is written as
\begin{align}
 V_{pq,rs} &= \sum_k^{N_\text{Chol}} R_{pq,k} R_{rs,k} & \text{for}\; pq, rs \in \mathcal{S}\,.
\end{align}
The number of Cholesky vectors $N_\text{Chol}$ is increased until the Cholesky vectors
matrix $\Matrix{R}$ satisfies the Coulomb matrix trace condition\cite{harbrecht2012low}
\begin{equation}
  \Tr\left(\Matrix{V}\right) - \Tr\left( \Matrix{R} \Matrix{R}^T \right) < \tau_\text{Chol}\,. \label{cholesky-condition}
\end{equation}
Unlike density fitting, the Cholesky decomposition of \Matrix{V} automatically achieves
an arbitrary precision level within the machine limits, without the need for
supplying a predefined auxiliary basis set.
That is in line with the requirement for extra numerical precision for $n$-body
noncovalent energies.\cite{richard2014understanding}
In practice, we apply Eq.~\ref{cholesky-condition} with a wide margin of safety
to ensure that the Coulomb matrix decomposition does not contribute to
the overall numerical error.
(See Table~\ref{numerical-thresholds} for the threshold values.)
As a consequence our computations involve the number of Cholesky vectors
which is significantly larger than in typical calculations involving
the decomposition of the Coulomb integrals.
For example, for water dimer in equilibrium geometry, the number of 
Cholesky vectors is six times the number of atomic orbitals. 
However, this does not affect significantly the cost of the RPA program
as the Cholesky vectors are further contracted into a much more compact basis spanning
the dominant eigenspace of \Matrix{\Pi}, as discussed later in the text.

We employ the Cholesky algorithm described in  Ref.~\citenum{aquilante2011choleski}
with the following modifications: \begin{inparaenum}[(i)] \item all AO pair indices 
belong to the set $\mathcal{S}$ defined in Eq.~\ref{significant-set},
\item the convergence condition involves the trace error of the Coulomb matrix.
\end{inparaenum}{}In contrast to the usual condition of the minimum decomposed diagonal
element, the condition given in Eq.~\ref{cholesky-condition} leads to extra numerical precision
in absolute energies, and thus less reliance on error cancellation for energy differences.

Having the Cholesky-decomposed Coulomb matrix, we now use it
to rewrite the expression for the correlation energy.
To this end we consider the lowest order term in Eq.~\ref{rpa-energy}.
In matrix form one can write
\begin{multline}
  \Tr(\chi^0(u) \upsilon_\text{ee}) = \sum_{pq,rs \in \mathcal{S}} \chi^0_{pq,rs}(u) V_{rs,pq} \\
  = \Tr\left(\Matrix{\chi}^0(u) \Matrix{R} \Matrix{R}^T\right)
  = \Tr\left(\Matrix{R}^T \Matrix{\chi}^0(u) \Matrix{R}\right)\,. \label{chol-deriv}
\end{multline}
The formula is analogous to the one derived by Ren et~al.\cite{ren2012resolution},
except for the use of the Cholesky vectors instead of a one-center auxiliary basis set. 
We use the last expression of Eq.~\ref{chol-deriv} to define the auxiliary
matrix \Matrix{\Pi} in the Cholesky basis
\begin{equation}
  \Matrix{\Pi}(u) = -\Matrix{R}^T \Matrix{\chi}^0(u) \Matrix{R}\,.
\end{equation}
One can see that the dimension of $\Matrix{\Pi}(u)$ is given by the number of Cholesky 
vectors $N_\text{Chol} \sim \mathcal{O}(N)$ instead of the usual $N_\text{occ} N_\text{virt}$.
Our RPA program stores and computes only the matrix elements corresponding
to $p \ge q$ and $pq \in \mathcal{S}$. 
We also halve the cost of the dominant step, i.e., the matrix
multiplication $\Matrix{\chi}^0(u) \Matrix{R}$,
by utilizing the permutational symmetry $\chi^0_{pq,rs}=\chi^0_{rs,pq}$. 

At this point one could diagonalize or LU-decompose $\Matrix{\Pi}(u)$ to obtain its eigenvalues
and hence the correlation energy.
Remarkably, a further reduction of the computational cost is possible at a given precision level without additional
assumptions on the sparsity of $\rho^\text{occ}$ and $\rho^\text{virt}$. 
As shown by Galli et~al.,\cite{wilson2008efficient,lu2008dielectric,wilson2009iterative} the RPA dielectric
matrix\cite{hybertsen1987ab} $\epsilon_\text{RPA}=1-\upsilon_\text{ee} \chi^0$, which is closely related to \Matrix{\Pi}
defined in this work, can be accurately reconstructed from
a small subset of its most heavily screened eigenpotentials. 
That idea has been employed by Nguyen and de Gironcoli\cite{nguyen2009efficient} to compute the RPA correlation energy
using the eigenpotentials of $\chi^0$ obtained with first-order density-functional perturbation
theory (DFPT).\cite{baroni2001phonons} 
In their plane-wave/pseudopotential code,\cite{giannozzi2017advanced} which scales
as $\mathcal{O}(N^4)$, the authors of Ref.~\citenum{nguyen2009efficient} solve the perturbed
Kohn-Sham equations for the density perturbation $\delta \rho$, which is the result of $\chi^0$ acting
on a trial eigenpotential. 
A repeated insertion of trial potentials into the first-order DFPT equations refines the guess eigenvectors
of $\chi^0$ for each frequency of the ACFD integral, without a summation over virtual states.
In a recent work, Hellgren et al.\cite{hellgren2018beyond} have demonstrated
the viability of that approach for RPAX calculations for noncovalent dimers.
(See also Ref.~\citenum{govoni2015large} for details of the projective eigendecomposition
of the dielectric screening method in the GW calculations.) 

Here, we introduce an AO basis method of computing $E_\text{c}$ by projecting
$\Matrix{\Pi}$ onto its dominant eigenvectors. 
The novelty of our approach lies in combining the eigendecomposition of $\Matrix{\Pi}$ 
with a cubic scaling AO method based on the Cholesky decomposition. 

Let the eigenvalues of the positive-definite matrix $\Matrix{\Pi}(u)$ be
\begin{equation}
  \lambda_1 \ge \lambda_2 \ge \ldots \ge \lambda_{N_\text{Chol}} \ge 0\,.
\end{equation}
We take a subset of $N_\text{eig}$, with $N_\text{eig} \le N_\text{Chol}$, largest eigenvalues 
$\lambda_1, \lambda_2, \ldots, \lambda_{N_\text{eig}}$
and store them in a matrix $\Matrix{G}$.
The matrix $\Matrix{G}$ then transforms $\Matrix{\Pi}$ to its effective reduced-dimension
form defined as
\begin{equation}
  \Matrix{\Pi}' = \Matrix{G}^T \Matrix{\Pi} \Matrix{G}\, .
\end{equation}
Given a user-defined threshold $\tau_\text{trace}$, 
we adjust $N_\text{eig}$ by appending $\Matrix{G}$ with new columns until the difference
between the exact and the reduced-dimension traces satisfies
\begin{equation}
 0 \le \Tr(\Matrix{\Pi}) - \Tr\left( \Matrix{\Pi}' \right) = \sum_{k=N_\text{eig}+1}^{N_\text{Chol}} \lambda_k < \tau_\text{trace}
\end{equation}
Importantly, the trace error in $\Matrix{\Pi}'$ introduces a quadratic error per single frequency in the ACFD integral
\begin{multline}
  \delta = \Tr(\ln(\Matrix{1}+\Matrix{\Pi})) - \Tr\left(\Matrix{\Pi}\right) \\
  - \left(\Tr(\ln(\Matrix{1}+\Matrix{\Pi}') - \Tr\left(\Matrix{\Pi}'\right) \right) \\
         = \left| \sum_{k=N_\text{eig}+1}^{N_\text{Chol}} \ln(1 + \lambda_k) - \lambda_k \right| 
  \le \left| \sum_{k=N_\text{eig}+1}^{N_\text{Chol}} \frac 1 2 \lambda_k^2 \right | \\
  \le \frac 1 2 \left| \sum_{k=N_\text{eig}+1}^{N_\text{Chol}} \lambda_k \right|^2 < \frac 1 2 \tau_\text{trace}^2 \,.
\label{quadratic-error}
\end{multline}
The RPA correlation energy expressed with the effective matrix \Matrix{\Pi}$'$ is
\begin{equation}
  E_\text{c} = \frac{1}{2 \pi} \int_0^\infty 
  \Tr \left( \ln(1 + \Matrix{\Pi}'(u)) - \Matrix{\Pi}'(u)  \right) \, \D u \,.\label{effective-acfd-formula} 
\end{equation}
The trace error $\delta$ is summed up over about 10-20 points on the frequency grid. 
Nonetheless, as the energy contributions fall off steeply 
with increasing frequency, the trace error at a few lowest frequencies dominates the integrated error.

Of course there would be no computational gain if the vectors of $\Matrix{G}$ were calculated by exact 
diagonalization for $\Matrix{\Pi}(u)$ at every frequency point.
To solve the above issue, we compute $\Matrix{G}$ only once, for the lowest frequency of the numerical grid
and use it for all other frequencies.
To avoid the diagonalization of the full-dimension matrix $\Matrix{\Pi}$, 
we build the matrix $\Matrix{G}$ by applying a variant of the subspace iteration method starting from random guess vectors, as
presented by Saibaba et al.\cite{saibaba2017randomized} 
Given the matrix $\Matrix{\Pi}$ for the lowest frequency, we carry out the iterations
\begin{align}
  \Matrix{G}^{[0]} &\leftarrow \text{random numbers from $\mathcal{N}(0,1)$}\\
  \Matrix{G}^{[m]} &\leftarrow \text{QR decomposition of $\Matrix{\Pi} \Matrix{G}^{[m-1]}$} \,.
\end{align}
As shown in Ref.~\citenum{saibaba2017randomized}, the trace error resulting from the subspace iteration method
approaches the sum of neglected small eigenvalues exponentially fast with the number of iterations, which makes
valid our error bound of Eq.~\ref{quadratic-error}. 
In all RPA calculations we used $m=2$ iterations to obtain $\Matrix{G}$. 
We observed only an insignificant difference in the number of basis vectors
as compared to exact diagonalization. 
For water dimer in equilibrium configuration, the exact diagonalization
yields $598$ effective basis vectors, whereas with the subspace iteration approach the number is $603$.

  The complete computational scheme involves both dielectric eigenvectors, i.e., the eigenvectors of \Matrix{\Pi},
  and the Cholesky decomposition of the Coulomb matrix. To reduce the number of floating-point operations
  and storage requirements, we obtain $\Matrix{G}$ for the full-dimension $\Matrix{\Pi}$ at the lowest frequency
  and use it to build a new matrix
  \begin{equation}
    \Matrix{R}' = \Matrix{R} \Matrix{G}
  \end{equation}
  At this point we deallocate \Matrix{R} as it is no longer needed. We subsequently
  reuse $\Matrix{R}'$ to obtain the auxiliary matrix
  \begin{equation}
    \Matrix{\Pi}'(u) = -\Matrix{R}'^T \Matrix{\chi^0}(u) \Matrix{R}'
  \end{equation}
  at all frequencies. With the screening condition of Eq.~\ref{significant-set}, the dimension
  of $\Matrix{R}'$ is $N_\text{AO}\times N_\text{eig}$.
  The most compute-intensive steps are the formation of $\Matrix{\chi^0}(u) \Matrix{R}'$
  (linear speed-up when using the dominant eigenspace) and the matrix
  multiplication $\Matrix{R}'^T \left(\Matrix{\chi^0}(u) \Matrix{R}'\right)$ (quadratic speed-up).
  In parallel computations, the blocks of $\Matrix{R}'$ are distributed between concurrent
  processes and each process builds its own chunk of $\Matrix{\chi^0}(u) \Matrix{R}'$.
  
  To get properly size-extensive interaction energy, we use the same
  effective eigenspace for the bound complex and for all its subsystems.
  For example, when calculating nonadditive three-body noncovalent interaction
  energies, we obtain the matrix $\Matrix{R}'$ for a trimer $ABC$ and reuse it
  for dimers $AB$, $BC$, $AC$, as well as for monomers $A$, $B$, and $C$.

\subsection{Numerical integration} \label{quadrature-description}

In our implementation there are two integrals which need to be calculated numerically, 
similar to the implementation of Kaltak et al.\cite{kaltak2014low,kaltak2014cubic}
First, the correlation energy is obtained by integration over an 
imaginary frequency grid.
Second, an imaginary time grid is needed to represent Green's functions from which
the response function is calculated via the Laplace transform.

The frequency integral of Eq.~\ref{rpa-energy} is approximated as
\begin{equation}
  E_\text{c} = \frac{1}{2 \pi} \sum_{k=1}^{n} w_{kn} \Tr\left[
  \ln(\Matrix{1} + \Matrix{\Pi}'(u_{kn})) - \Matrix{\Pi}'(u_{kn})
    \right]\,,
\end{equation}
where the frequencies $u_{kn}$ and weights $w_{kn}$ are given by the points $x_{kn}^\text{GL}$ 
and weights $w^\text{GL}_{kn}$ of an $n$-point Gauss-Legendre quadrature
mapped onto the half-infinite interval
\begin{align}
  u_{kn}(\zeta) &= \zeta \frac{1 + x^\text{GL}_{kn}}{1 - x^\text{GL}_{kn}} \\
  w_{kn}(\zeta) &= \zeta \frac{2 w^\text{GL}_{kn}}{(1 - x^\text{GL}_{kn})^2}\,.
\end{align}
Moreover, we have modified the mapping used by Ren et al.\cite{ren2012resolution} by making
the parameter $\zeta$ a variable adjusted per system.
To find the optimal $\zeta$ we consider the frequency-dependent part of Eq.~\ref{real-chi}
as a test function.
The exact integral of the test function and its quadrature approximation for a number of nodes $n$
are given as 
\begin{align}
  \mathcal{I}_\text{exact}(d_{ai}) = \int_0^\infty \frac{\D\,u}{d^2_{ai}+u^2} = \frac{\pi}{2 d_{ai}} \\
  \mathcal{I}_\text{quad}(d_{ai};\zeta,n) = \sum_{k=1}^{n} \frac{w_{kn}(\zeta)}{d^2_{ai}+u_{kn}(\zeta)^2}\,.
\end{align}
The optimal $\zeta$ then minimizes the squared error averaged over the distribution
of orbital energy differences 
\begin{multline}
  \zeta(n) = \argmin_{\zeta'} \max_\mu
  \sum_\nu h(d_{\nu \mu}) \left(\mathcal{I}_\text{exact}(d_{\nu \mu}) \right.
  \\ \left. - \mathcal{I}_\text{quad}(d_{\nu \mu};\zeta',n)\right)^2 \,. \label{zeta-def}
\end{multline}
The weighting function $h(d_{\nu \mu})$ counts how many orbital energy differences
fall into a histogram bin $\nu$ for a system $\mu$. 
The set of all energy differences is divided into $100$ bins for each system. 
In the case of a trimer, the maximum in Eq.~\ref{zeta-def} is taken over the set 
of the trimer and all of the dimer and monomer subsystems
so that the interacting noncovalent complex and all its subsystems share the quadrature points and weights. 
Owing to the common grid, the interaction energy includes only the physical interactions and not the effects
of changing the grid between subsystems. 
In particular, at large separations, the interaction energy properly goes to zero. 
To be able to define the common grids the SCF for all subsystems must precede the RPA program.
The number of nodes $n$ is the smallest integer which satisfies the root-mean square error
and the maximum relative error conditions
\begin{multline}
    \max_\mu \left( \sum_\nu h(d_{\nu \mu}) \left(\mathcal{I}_\text{exact}(d_{\nu \mu}) \right. \right. \\
    \left. \left. - \mathcal{I}_\text{quad}(d_{\nu \mu};\zeta(n),n)\right)^2 \right)^{1/2} < \tau_\text{freq,RMSD}
\end{multline}
\begin{multline}
  \max_{\mu \nu} \left|\frac{\mathcal{I}_\text{exact}(d_{\nu \mu})-\mathcal{I}_\text{quad}(d_{\nu \mu};\zeta(n),n)}{\mathcal{I}_\text{exact}(d_{\nu \mu})}\right| \\
  < \tau_\text{freq,MaxRel}\,.
\end{multline}

The imaginary time integral includes an oscillatory integrand depending
on the frequency of the density response function
\begin{multline}
  \chi^0_{pq,rs}(\imag u_{kn}) \\
  \begin{aligned}
    &=-4 \int_0^\infty \cos(u_{kn} t) \rho_{pr}^\text{occ}(t) \rho_{qs}^\text{virt}(t) \, \D t \\
    &= -4 \sum_{l=1}^{n'} w'_{ln'} \rho_{pr}^\text{occ}(t_{ln'}) \rho_{qs}^\text{virt}(t_{ln'})\,.
    \end{aligned}
\end{multline}
The parameters of the quadrature, $t_{ln'}$, $w'_{ln'}$, and $n'$, are adjusted for each $u_{kn}$. For near-zero frequencies,
we use the minimax quadrature of Takatsuka~et~al.,\cite{takatsuka2008minimax} which is designed for
decomposing the zero-frequency denominators occurring in Laplace-transformed MP2.\cite{haser1992laplace,haser1993moller}
For higher frequencies, we use the robust double exponential quadrature of Ooura and Mori.\cite{ooura1999robust}
(See Eq.~4.2 in Ref.~\citenum{ooura1999robust}.) Using the test function and its quadrature approximation
\begin{align}
  \mathcal{I'}_\text{exact}(d_{ai}, u_{kn}) &= \frac{d_{ai}}{d_{ai}^2+u_{kn}^2} \\
  \mathcal{I'}_\text{quad}(d_{ai}, u_{kn};n') &= \sum_{l=1}^{n'} w'_{ln'}(u_{kn}) \exp\left(-d_{ai} t_{ln'}(u_{kn})\right)
\end{align}
we fix the number of grid points $n'$ as the smallest integer satisfying both
\begin{multline}
  \max_\mu \left( \sum_\nu h(d_{\nu \mu}) \left(\mathcal{I'}_\text{exact}(d_{\nu \mu}) \right.\right. \\
  \left.\left. - \mathcal{I'}_\text{quad}(d_{\nu \mu};u_{kn},n')\right)^2 \right)^{1/2}  < \tau_\text{imag,RMSD}
\end{multline}
and
\begin{multline}
  \max_{\mu \nu} \left|\frac{\mathcal{I'}_\text{exact}(d_{\nu \mu})-\mathcal{I'}_\text{quad}(d_{\nu \mu};u_{kn},n')}{\mathcal{I'}_\text{exact}(d_{\nu \mu})}\right| \\
  < \tau_\text{imag,MaxRel}\,.
\end{multline}
The transition point between the minimax and double exponential quadratures depends on which
approach is able to satisfy the above conditions with a smaller number of points. Similarly
to the frequency quadrature, the complex and all its subsystems share the same imaginary time
grid.

The thresholds controlling the grid accuracy as well as other thresholds used
for the RPA correlation energy are summarized in Table~\ref{numerical-thresholds}.
The specified values are hand-tuned against accurate RPA calculations on our
calibration set of noncovalent dimers and trimers. 
The data in Table~\ref{rpa-settings} demonstrate the influence of the numerical settings 
on 2-body and 3-body interaction energies.

\begin{table*}[tb]
  \setlength{\tabcolsep}{5pt}
  \captionsetup{width=0.75\textwidth}
  \caption{Thresholds, in atomic units, controlling the numerical precision of the RPA correlation energy.}
  \label{numerical-thresholds}
  \begin{tabular}{llll}
    \toprule
                   & $1$           &  $2$              &         $3$ \\
    \midrule
    $\tau_\text{screen}$      & $10^{-6}$          & $10^{-7}$          &  $10^{-8}$ \\
    $\tau_\text{Chol}$        & $10^{-2}$          & $10^{-3}$          &  $10^{-4}$ \\
    $\tau_\text{freq,RMSD}$   & $10^{-6}$          & $10^{-6}$          &  $10^{-6}$ \\
    $\tau_\text{freq,MaxRel}$ & $10^{-3}$          & $10^{-3}$          &  $10^{-3}$ \\
    $\tau_\text{imag,RMSD}$   & $10^{-6}$          & $10^{-6}$          &  $10^{-6}$ \\
    $\tau_\text{imag,MaxRel}$ & $10^{-3}$          & $10^{-3}$          &  $10^{-3}$ \\
    $\tau_\text{trace}$       & $\sqrt{10^{-5}}$   & $\sqrt{10^{-7}}$   &  $10^{-5}$ \\
    \bottomrule
  \end{tabular}
\end{table*}

\begin{table*}[tb]
  \setlength{\tabcolsep}{6pt}
  \captionsetup{width=0.75\textwidth}
  \caption{Numerical precision of the RPA correlation energies. 
    The uppermost row contains the reference correlation
    parts of two- and three-body energies.
    The rows corresponding
    to the parameter sets 1\ldots3 (defined in Table~\ref{numerical-thresholds}) contain
    the numerical errors with respect to the reference.
    All energies are in kcal/mol.
    The aug-cc-pVQZ basis is used for all datapoints.
    The spacing between atoms in the linear configuration is $R=\SI{3}{\AA}$.
    The neon trimer at $\ang{26.6}$ is an isosceles triangle with the base $R=\SI{2.8}{\AA}$.}
  \label{rpa-settings}
  \begin{tabular}{lrrrrr}
    \toprule
    Numerical precision   & \ce{Ne2}           &     \ce{(NH3)2}       &    linear \ce{Ne3}   &    \ce{(H2O)2}         & \ce{Ne3} ($\ang{26.6}$)   \\
    \midrule
    Reference    & $-0.167650$          &     $-1.783236$         &     $-0.0003748$      &    $-1.320970$         &     $0.322301$    \\
    $1$          & $-4\times 10^{-5}$             &     $-3\times 10^{-5}$            &     $-3\times 10^{-7}$                &    $-5\times 10^{-4}$ & $-1 \times 10^{-4}$ \\
    $2$          & $-2\times 10^{-5}$             &     $3\times 10^{-5}$             &     $1\times 10^{-7}$                 &    $3\times 10^{-5}$  & $3 \times 10^{-6}$   \\
    $3$          & $-6\times 10^{-6}$             &     $-5\times 10^{-6}$            &     $4\times 10^{-8}$                 &    $-1\times 10^{-5}$ & $4 \times 10^{-6}$  \\
    \bottomrule
  \end{tabular}
\end{table*}

\subsection{Singles correction}

From the point of view of ordinary Rayleigh-Schrodinger perturbation theory,
the Kohn-Sham orbitals are noncanonical and the correlation energy includes
nonzero terms related to single excitations from the Kohn-Sham determinant.
The usefulness of a beyond-RPA approach including singles was first demonstrated
by Ren et al.,\cite{ren2011beyond} who, in their study of noncovalent interaction
energies, employed the Hartree-Fock (HF) orbitals to compute the mean-field part of the RPA
total energy and the Kohn-Sham orbitals to obtain the correlation part.
The renormalized singles formula which sums up singles through infinite order
was later derived in Ref.~\citenum{ren2013renormalized}.
Finally, Klimes et al.\cite{klimes2015singles} derived the renormalized singles term 
in a form that is applied in this work
\begin{multline}
  E_\text{c}^\text{RSE} = 2 \Tr\left( \Matrix{\rho}^\text{HF} \Matrix{F}^\text{HF}\left[\Matrix{\rho}^\text{DFT}\right] \right) \\
  - 2 \Tr\left( \Matrix{\rho}^\text{DFT} \Matrix{F}^\text{HF}\left[\Matrix{\rho}^\text{DFT} \right] \right)\,.
\label{singles-correction}
\end{multline}
The self-consistent Kohn-Sham orbitals are used to build the density matrix $\Matrix{\rho}^\text{DFT}$ and
the HF hamiltonian $\Matrix{F}^\text{HF}\left[\Matrix{\rho}^\text{DFT}\right]$. In contrast to the
approach presented in Ref.~\citenum{ren2011beyond}, here the density matrix $\Matrix{\rho}^\text{HF}$
is obtained in a single step diagonalization from the eigenvectors of $\Matrix{F}^\text{HF}\left[\Matrix{\rho}^\text{DFT}\right]$.
The computational cost of RSE equals that of a single Fock matrix evaluation.
A higher-accuracy variant of the singles correction includes the density from the $GW$ method
instead of $\Matrix{\rho}^\text{HF}$.\cite{klimes2015singles}

Adding the singles corrections has been shown to improve the RPA binding energies
for several systems.\cite{ren2011beyond,klimes2015singles}
However, how they affect three-body energies is unknown and this is one of our 
interests in this work.
Another point we try to understand is how do the singles depend on the input DFT orbitals.

\section{Numerical results}
\subsection{Technical details}
In this study we use RPA and wavefunction methods to obtain two-body interaction energies
\begin{equation}
 E_\text{int} = E(AB) - E(A) - E(B)
\end{equation}
and nonadditive three-body interaction energy components
\begin{multline}
  E_\text{int}[3,3] = E(ABC) - E(AB) - E(BC) \\ - E(AC)
  + E(A) + E(B) + E(C)
\end{multline}
In all calculations, the monomer, dimer, and trimer energies
are computed in the trimer basis set and at the same geometries
as in the trimer.

The RPA calculations were carried out using in-house software implementing the algorithms presented
in Sections~\ref{implementation-description} and~\ref{quadrature-description}.
The code also calculates the exact-exchange (EXX) component of the energy (the Hartree-Fock-like part of the RPA energy
evaluated with DFT orbitals) and the singles correction.
The Molpro package\cite{werner2012molpro} was used to obtain the HF energy and correlation energy at the coupled cluster
level and using different orders of M{\o}ller-Plesset perturbation theory.\cite{MOLPRO,hampel1992comparison}
Moreover, we used Molpro\cite{werner2012molpro} to obtain
three-body dispersion energies.
Finally, VASP was used to obtain a GWSE correction for neon trimer.\cite{kaltak2014low,kaltak2014cubic,klimes2015singles,kresse1996efficient}

The correlation energies, including RPA, typically converge slowly with the basis set size.
We therefore extrapolated all wavefunction and RPA correlation energies to complete basis-set (CBS) limit
$E_\text{CBS}$ using the two-point scheme of Halkier~{\it et al.}\cite{halkier1999basis}
\begin{equation}
E_\text{CBS}={(X+1)^3E_{X+1} - X^3 E_X\over (X+1)^3-X^3}
\end{equation}
here $X$ denote the cardinal number of a basis set and $E_X$ is the corresponding energy.
Dunning's correlation consistent basis sets are used throughout this work.\cite{schuchardt2007basis}
The extrapolation scheme for noble gases is aug-cc-pVQZ $\rightarrow$ aug-cc-pV5Z; for all other
systems the scheme used is aug-cc-pVTZ $\rightarrow$ aug-cc-pVQZ.
The DFT energies and HF components of the wavefunction and RPA energies were obtained 
using the basis set with the largest cardinal number used for a given system and were not extrapolated.

Four different exchange-correlation functionals were used to generate the orbital input for RPA, 
namely PBE\cite{perdew1996generalized} and its variant PBE0 including 25\% of EXX,\cite{adamo1999toward} 
the meta-GGA SCAN\cite{sun2015strongly} and SCAN0 which is its hybrid variant with 25\% of EXX.\cite{hui2016SCAN}
The label RPA($X$) corresponds to the RPA energy evaluated with orbitals and orbital energies 
of the DFT exchange-correlation model $X$. RPA($X$)+RSE denotes RPA with the addition of the renormalized
singles energy.\cite{klimes2015singles}

\subsection{Noble gas dimers}
\begin{figure*}[tb]
  \includegraphics[width=0.75\textwidth]{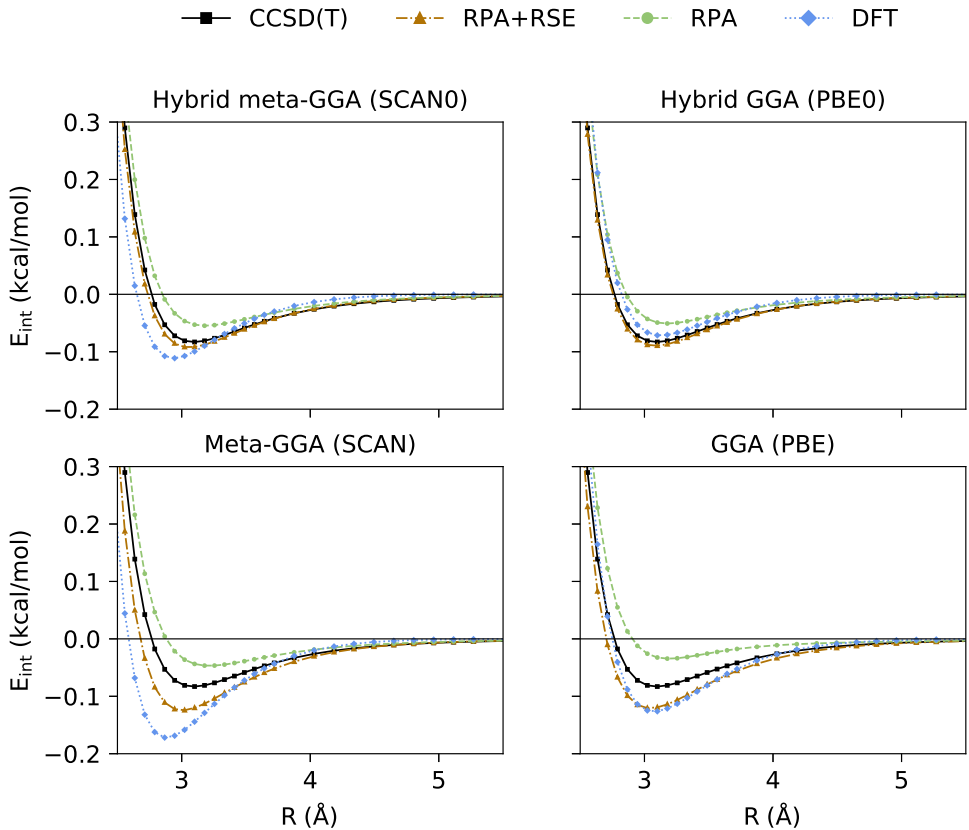}
  \caption{Neon dimer interaction energy computed using four sets of DFT orbitals.} \label{neon-dimer}
\end{figure*}

\begin{figure*}[tb]
  \includegraphics[width=0.75\textwidth]{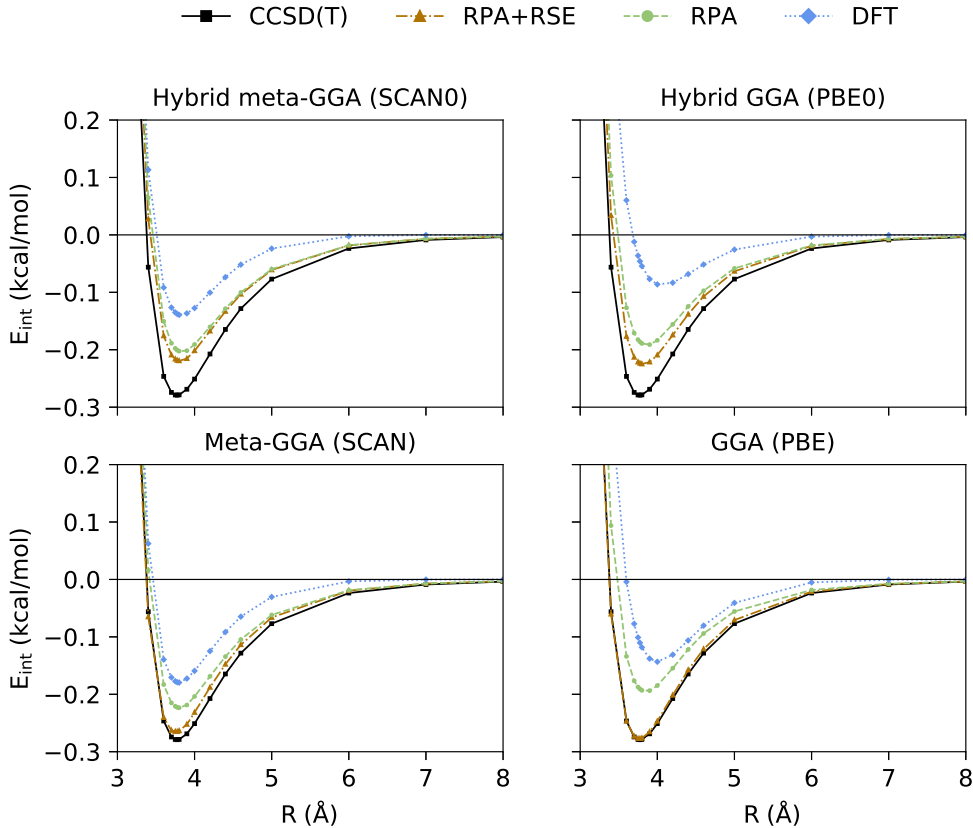}
  \caption{Argon dimer interaction energy computed using four sets of DFT orbitals.
    The reference coupled-cluster curve is taken from Ref.~\citenum{patkowski2005accurate}.} \label{argon-dimer}
\end{figure*}

\begin{figure*}[bt]
  \includegraphics[width=0.75\textwidth]{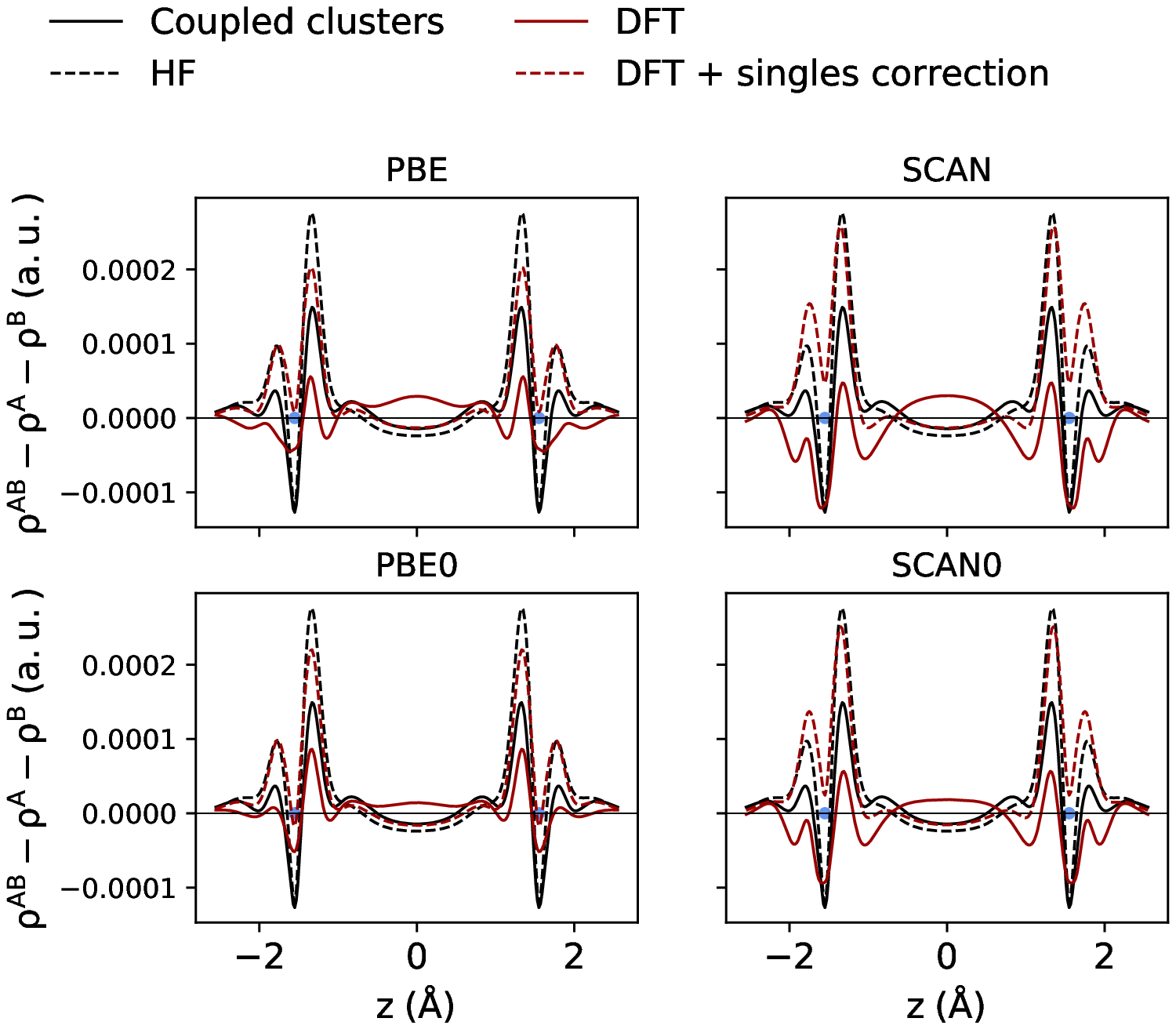}
  \caption{Change in DFT and coupled-cluster electron densities induced
    by the interaction in the Ne dimer at $R=3.1$~\AA.
    The atoms are placed at $z=\pm \SI{1.55}{\AA}$.
    The drawn density resides at
    the interval $AB$ between endpoints  $A=(0.125,0,-2.55)~\AA$ and $B=(0.125,0,+2.55)~\AA$, i.e.,
    on the line placed $0.125~\AA$ above the bond axis. The CC3-level coupled-clusters density is generated using
    the software described in Refs.~\citenum{tucholska2014transition} and~\citenum{tucholska2017transition}.
    Both coupled-cluster and DFT calculations employ the aug-cc-pCVQZ basis.} \label{Ne2Rho}
\end{figure*}

The vast majority of existing work on RPA for noncovalent systems is related to the interaction
energies of molecular dimers\cite{eshuis2011a,bleiziffer2013efficient} 
or adsorption energies.\cite{torres2017adsorption,alhamdani2017properties,brandenburg2019physisorption}
As our conclusions regarding three-body systems are best understood
in the context of two-body results, we briefly demonstrate the performance of RPA
for the dimers of neon and argon. 
In particular, we are interested in the comparison of DFT vs. RPA, the differences between 
RPA energies computed with different orbital sets, and the effect of the singles correction.

Semilocal and DFT functionals lack long-range correlation and as a consequence
the interaction energy decays too quickly in the tail region of both neon and argon.
As expected, RPA, which correctly accounts for the dispersion energy,
corrects for that deficiency for all orbital sets, see Figures~\ref{neon-dimer}
and~\ref{argon-dimer}.

At the equilibrium separation and in its vicinity, the DFT error
becomes more difficult to predict.
Two principal sources of error, the missing long-range dispersion
and artificial binding due to the exchange energy, determine the total deviation.
For neon, all DFT functionals except for PBE0 overbind, while for
argon all DFT methods underbind.

The variation of the RPA results with the change of the orbital set
is limited, but quantitatively important.
As evident from Figures~\ref{neon-dimer} and~\ref{argon-dimer},
RPA without RSE underestimates the curve depth around the equilibrium.
The RSE correction always improves the energy upon bare RPA for argon.
As expected, the RSE correction is smaller for the hybrid functionals.
For \ce{Ne2}, RPA+RSE using hybrid DFT orbitals is closer to the reference
than for the pure DFT input. 
In contrast, using pure DFT gives better results for \ce{Ar2}.

It is useful to compare the electron density predicted by DFT schemes with highly 
accurate density to identify the sources of errors and see the effect of the singles corrections.
To this end we plot the difference between the density
of the equilibrium Ne dimer at $R=\SI{3.1}{\AA}$ and the sum of isolated atom densities
for each of the four approximate functionals, as well as for HF and for the coupled cluster (CC)
schemes in Figure~\ref{Ne2Rho}.
The singles correction corresponds to the density obtained with a single HF iteration starting
from the converged DFT self-consistent field.
In the density difference plot, both HF and CC give charge depletion in the midpoint between the atoms
while all the DFT approximations used show charge accumulation.
While this artifact of approximate DFT could be partly caused by inaccurate correlation functional, 
inaccurate description of exchange is a more likely cause.
This is because the difference between the HF and CC curves around midpoint is much smaller 
than the difference between the HF data and result of any DFT functional.
Moreover, the incorrect accumulation is somewhat reduced when going from pure functionals to hybrids, 
also pointing to incorrect description of exchange.
Finally, the addition of the singles correction has the largest effect, it completely removes the 
artifact and the density difference becomes closer to that of HF or CC.

\subsection{The 3B-69 test set}

\begin{table*}[tb]
  \caption{Average errors (kcal/mol) for the 3B-69 set of trimers.\cite{rezac2015benchmark}
    Data for the M{\o}ller-Plesset perturbation theory approximations are taken from Ref.~\citenum{rezac2015benchmark}.}
  \label{rezac-errors}
    \setlength{\tabcolsep}{12pt}
  \begin{tabular}{lrrr}
    \toprule
    Method & MSE & MUE & RMSE \\
    \midrule
    RPA(SCAN0)     & $-$0.017 & 0.018 & 0.023 \\
    RPA(SCAN0)+RSE & $-$0.027 & 0.028 & 0.034 \\
    RPA(PBE0)      & $-$0.026 & 0.026 & 0.033 \\
    RPA(PBE0)+RSE  & $-$0.020 & 0.023 & 0.029 \\
    RPA(PBE)       & $-$0.041 & 0.044 & 0.054 \\
    RPA(PBE)+RSE   & $-$0.013 & 0.026 & 0.038 \\
    MP2            & $-$0.039 & 0.045 & 0.059 \\
    MP3            & 0.022 & 0.026 & 0.035 \\
    MP2.5          & $-$0.009 & 0.014 & 0.019 \\
    SCAN0          & $-$0.054 & 0.065 & 0.081 \\
    PBE0           & 0.017 & 0.039 & 0.053 \\
    PBE            & 0.068 & 0.093 & 0.116 \\
    \bottomrule
  \end{tabular}
\end{table*}

\begin{table*}[tb]
  \caption{Average errors (kcal/mol) for the low, medium, and high dispersion subsets of the 3B-69 set of trimers
 as defined by \v{R}ez\'{a}\v{c} and co-workers.\cite{rezac2015benchmark}
    Data for the M{\o}ller-Plesset perturbation theory approximations are taken from Ref.~\citenum{rezac2015benchmark}.
}
  \label{rezac-errors:LMH}
    \setlength{\tabcolsep}{5pt}
  \begin{tabular}{lrrrrrrrrr}
    \toprule
     & \multicolumn{3}{c}{Low dispersion} & \multicolumn{3}{c}{Medium dispersion} & \multicolumn{3}{c}{High dispersion} \\
    Method & MSE & MUE & RMSE& MSE & MUE & RMSE & MSE & MUE & RMSE \\
    \midrule
    RPA(SCAN0)     & $-$0.013 & 0.014& 0.019 & $-$0.019 &0.019 & 0.024& $-$0.022 & 0.023&0.026\\
    RPA(SCAN0)+RSE & $-$0.020 & 0.021& 0.027 & $-$0.029 &0.029 & 0.033& $-$0.035 & 0.036&0.041\\
    RPA(PBE0)      & $-$0.019 & 0.020& 0.028 & $-$0.027 &0.027 & 0.031& $-$0.034 & 0.034&0.038\\
    RPA(PBE0)+RSE  & $-$0.018 & 0.023& 0.031 & $-$0.019 &0.020 & 0.024& $-$0.023 & 0.026&0.029\\
    RPA(PBE)       & $-$0.039 & 0.044& 0.057 & $-$0.036 &0.040 & 0.049& $-$0.049 & 0.049&0.055\\
    RPA(PBE)+RSE   & $-$0.025 & 0.041& 0.054 & $-$0.005 &0.016 & 0.022& $-$0.006 & 0.018&0.021\\
    MP2            & $-$0.015 & 0.027& 0.039 & $-$0.044 &0.048 & 0.061& $-$0.064 & 0.066&0.074\\
    MP3            & 0.012    & 0.021& 0.026 & 0.018    &0.022 & 0.031& 0.038    & 0.038&0.046\\
    MP2.5          & $-$0.002 & 0.012& 0.016 & $-$0.013 &0.016 & 0.022& $-$0.013 & 0.015&0.020\\
    SCAN0          & $-$0.028 & 0.052& 0.069 & $-$0.061 &0.061 & 0.074& $-$0.082 & 0.085&0.099\\
    PBE0           & 0.000    & 0.051& 0.064 & 0.035    &0.039 & 0.052& 0.022    & 0.025&0.033\\
    PBE            & 0.021    & 0.084& 0.112 & 0.097    &0.098 & 0.121& 0.100    & 0.100&0.112\\
    \bottomrule
  \end{tabular}
\end{table*}

\begin{table*}[bt]
  \caption{Nonadditive energies for the \emph{challenging} and \emph{easy} subsets of the 3B-69 
    dataset. The subsets are specified according to the relative error of the RPA(SCAN0)
    nonadditive energy (explained in the main text). Energies are in kcal/mol. The reference, MP2, and MP3 energies
    are taken from Ref.~\citenum{rezac2015benchmark}.
    Here, $E_\text{disp}$ denotes the uncoupled\cite{misquitta2005intermolecular} three-body dispersion energy extrapolated
    using the aug-cc-pVTZ and aug-cc-pVQZ basis sets. RPA and RPA+RSE employ the SCAN0 orbitals.
  } \label{challenging-subset}
  \setlength{\tabcolsep}{5pt}
  \begin{tabular}{lrrrrrr}
    \toprule
    System                    & Ref. & $E_\text{disp}$ & MP2   & MP3 & RPA & RPA+RSE \\
    \midrule
    \multicolumn{7}{c}{Challenging subset} \\
    \midrule
    \ce{(CH3OH)2-ethyne} (3c) & 0.023   & 0.037           & $-$0.003 & 0.089 & 0.008   & $-$0.002 \\
    pyrazole (12b)            & 0.067   & 0.129           & 0.010 & 0.114 & 0.026    & 0.005 \\    
    triazine (13a)            & $-$0.005  & 0.008           & 0.013 & 0.002 & $-$0.011   & $-$0.010 \\
    succinic anhydride (18b)  & $-$0.001  & $-$0.003          & 0.003 & 0.001 & 0.007    & 0.004 \\ 
    benzene (19a)             & 0.048   & 0.204           & $-$0.054 & 0.126 & 0.021   & $-$0.012 \\ 
    benzene (19c)             & $-$0.027  & 0.085           & $-$0.061 & 0.016 & $-$0.044  & $-$0.053 \\
    p-benzoquinone (21b)      & 0.004   & 0.058           & $-$0.038 & 0.006 & $-$0.012  & $-$0.028 \\
    uracil (22a)              & $-$0.004  & 0.068           & $-$0.033 & 0.006 & $-$0.010  & 0.005 \\
    cyclobutylfuran (23a)     & 0.081   & 0.274           & $-$0.049 & 0.186 & 0.031   & $-$0.009 \\ 
    \midrule
    \multicolumn{7}{c}{Easy subset} \\
    \midrule
    water (1c)                & $-$2.416  & 0.068           & $-$2.472 & $-$2.404 & $-$2.411 & $-$2.461 \\
    acetonitrile (4c)         & $-$0.166  & $-$0.005          & $-$0.155 & $-$0.132 & $-$0.165 & $-$0.168 \\
    nitromethane (5c)         & 0.220   & $-$0.007          & 0.216 & 0.229 & 0.220    & 0.217 \\
    acetic acid (6a)          & 0.542   & 0.031           & 0.523 & 0.558 & 0.541    & 0.544 \\
    oxalic acid (7b)          & $-$1.198  & 0.012           & $-$1.199 & $-$1.170 & $-$1.201 & $-$1.228 \\
    acetamide (9c)            & $-$0.860  & $-$0.003          & $-$0.850 & $-$0.869 & $-$0.858 & $-$0.863 \\
    imidazole (10c)           & $-$1.636  & $-$0.013          & $-$1.631 & $-$1.608 & $-$1.628 & $-$1.665 \\
    maleic acid (20b)         & $-$1.449  & $-$0.004          & $-$1.428 & $-$1.419 & $-$1.451 & $-$1.496 \\
    p-benzoquinone (21c)      & 0.096   & 0.090           & 0.039 & 0.126 & 0.096    & 0.070 \\
    \bottomrule
  \end{tabular}
\end{table*}

\begin{figure*}[tb]
  \begin{center}
    \includegraphics[width=0.75\textwidth]{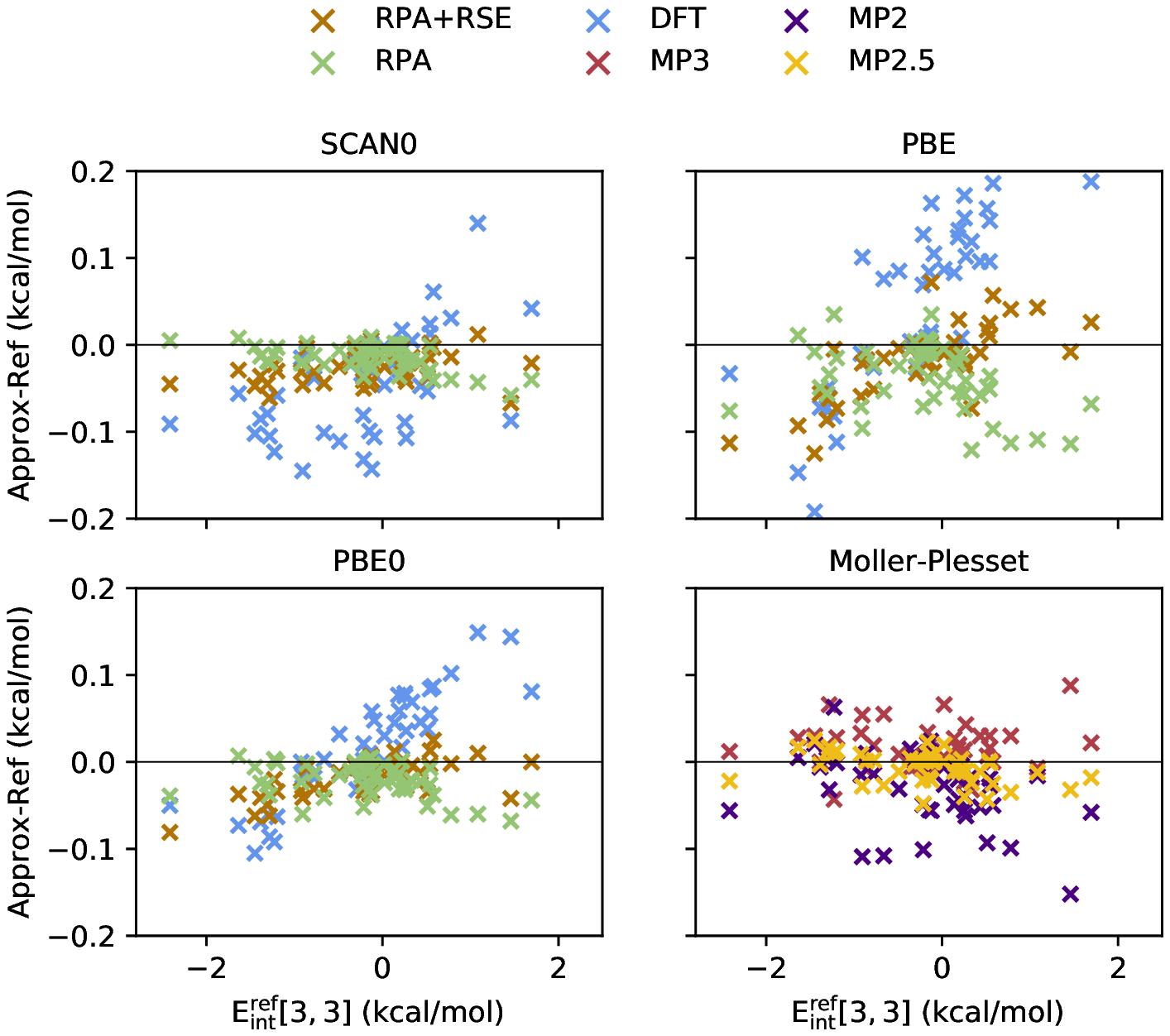}     
  \end{center}
  \caption{Signed errors of approximate methods vs the reference nonadditive interaction energy (kcal/mol)
    on the 3B-69 test set.\cite{rezac2015benchmark} } \label{fig:rezac}
\end{figure*}

We now turn to discuss the RPA results for predicting three-body energies
in the 3B-69 test set of Řezáč and coworkers.\cite{rezac2015benchmark}
The test set includes trimers of molecules composed of main-group elements. 
The reference energies employed in this work are taken
from Ref.~\citenum{rezac2015benchmark}.
The dataset avoids some of statistical biases by including
a mix of systems which interact with a varying amount of nonadditive dispersion. 

First we consider the RPA variants without the RSE correction.
The mean signed errors (MSE), mean unsigned errors (MUE),
and root-mean-square errors (RMSE) for RPA run with PBE, PBE0, and SCAN0 inputs
are summarized in Table~\ref{rezac-errors}.
The three-body nonadditive contributions improve when
going from the simplest to the most advanced exchange-correlation model.
Specifically, the RMSEs decrease from 0.054~kcal/mol for RPA(PBE),
over 0.033~kcal/mol for RPA(PBE0), to 0.023~kcal/mol for RPA(SCAN0).
For all aforementioned variants of RPA the MSE are negative, meaning
that the three-body energies are too attractive compared to the reference, 
see also Figure~\ref{fig:rezac}.

Without the presence of the RSE correction, we observe stark differences in
RPA's performance across the low, medium, and high dispersion subsets of the 3B-69 dataset. 
For RPA(PBE0) and RPA(SCAN0), the average errors increase with the fraction of the dispersion
energy component (Table~\ref{rezac-errors:LMH}).
While the above observation may look trivial for wavefunction based methods,
it is not obvious for DFT based schemes, including RPA.
As an example, the performance of RPA(PBE) cannot be rationalized
 in a simple way. 
Due to the sources of error inherited from the PBE orbitals,
 e.g., in the exchange and polarization nonadditivities,
the error distribution of RPA(PBE) is much more uniform across the subsets of 3B-69 compared
 to the other RPA variants.

We now turn to the results obtained for RPA with the RSE correction added.  
While RSE improved the accuracy of the RPA interaction energies for noble gas dimers
regardless of the orbital set, this is no longer the case for the 3B-69 dataset.
RPA(SCAN0)+RSE is nearly always worse than RPA(SCAN0).  
This reduced accuracy occurs for all the subsets, irrespective of the importance of dispersion,
see Table~\ref{rezac-errors:LMH}.
In this case, the RSE correction might overcorrect the errors in the three-body energies and terms beyond 
RSE are probably required to improve the accuracy of RPA(SCAN0).
For RPA(PBE0), the statistical errors decrease by 10 to 20~\% upon the addition of RSE.
Here the largest improvement occurs in the high and medium dispersion subsets (Table~\ref{rezac-errors}).
For the low dispersion subset, the errors increase for large negative three body energies
and decrease for the positive ones.
Typically, these correspond to systems with cooperative hydrogen bonds in the former case,
and a hydrogen bonded dimer and a spectator molecule not taking part in the hydrogen bonding in the latter case.
For RPA(PBE), we observe the same behavior of the RSE correction as for RPA(PBE0), 
only the reduction of errors brought by RSE is larger.

To establish the cost to accuracy ratio for RPA, we compare it to traditional post-HF approaches.
The worst RPA variant, RPA(PBE) is comparable in performance to MP2.
The best performing RPA variants, that is, RPA(SCAN0) and RPA(PBE0)+RSE, are more accurate
than MP2 and even better than MP3.  
This is a remarkable result considering the fact that MP3 scales with the sixth power of the system size,
while the computational cost of RPA increases only with the third power.
In fact, RPA(SCAN0) leads to an overall MUE of 0.018~kcal/mol, this is only 30~\% larger than
the MUE of 0.014~kcal/mol found for the MP2.5 approach and for the CCSD scheme.\cite{rezac2015benchmark}

We now attempt to  gain additional insight into the performance of the best performing variant, RPA(SCAN0).
We identify the systems 
for which RPA(SCAN0) exhibits exceptionally large and exceptionally small errors, i.e.,
the {\it challenging} and {\it easy} subsets according to the relative deviations of $E_\text{int}[3,3]$
from the reference.
The first nine systems with the errors above 50~\% are shown in Table~\ref{challenging-subset} as the 
{\it challenging} subset; the nine trimers with errors below 0.5~\% are shown as the {\it easy} subset.
The challenging systems, e.g., the trimers of benzene and uracil, are characterized
by two features: \begin{inparaenum}[(i)] \item the share of the third-order contributions
  to $E_\text{int}[3,3]$ is large, that is,
  \begin{equation}
    \left| \frac{E_\text{int}[3,3](\text{MP3})-E_\text{int}[3,3](\text{MP2})}{E_\text{int}[3,3](\text{CCSD(T)})} \right|  > 1
  \end{equation}
and \item the magnitude of the total nonadditive interaction energy is small. \end{inparaenum}
In those cases the RPA's errors are more apparent than in induction-dominated systems, where the
total nonadditive interaction is generally stronger.

Most of the easy systems, e.g., trimers of water and acetic acid, are polar.
The share of third-order M{\o}ller-Plesset contributions in $E_\text{int}[3,3]$
is small, and the magnitude of $E_\text{int}[3,3]$ is large. 
Alternatively, the three-body dispersion contribution in those systems is small 
compared to the total interaction (Table~\ref{challenging-subset}).

Let us briefly discuss the relation between the errors of the DFT functionals and the errors
of RPA used to perform the subsequent calculations.
First, one can notice that the results of PBE0 are, in terms of statistics, better than the 
results of RPA(PBE), see Table~\ref{rezac-errors:LMH}.
Even more, for the high dispersion subset, the MSE and MUE of PBE0 are on par with 
those of RPA(SCAN0), the best RPA scheme tested here.
Specifically, the error of PBE0 for the last challenging trimer, cyclobutylfuran 23a, is only $-0.005$~kcal/mol
while RPA(PBE0) differs by $-0.061$~kcal/mol from the reference.
However, the good performance for PBE0 is a result of cancellation of errors between
lack of long range correlation and spurious exchange binding.
Moreover, the (T) terms, not accounted for in RPA, and amounting to 0.039~kcal/mol for the cyclobutylfuran 23a
trimer, could represent a part of the RPA error in this and similar cases.

\subsection{Noble gas trimers}

\begin{figure}[tb]
  \includegraphics[width=0.5\textwidth]{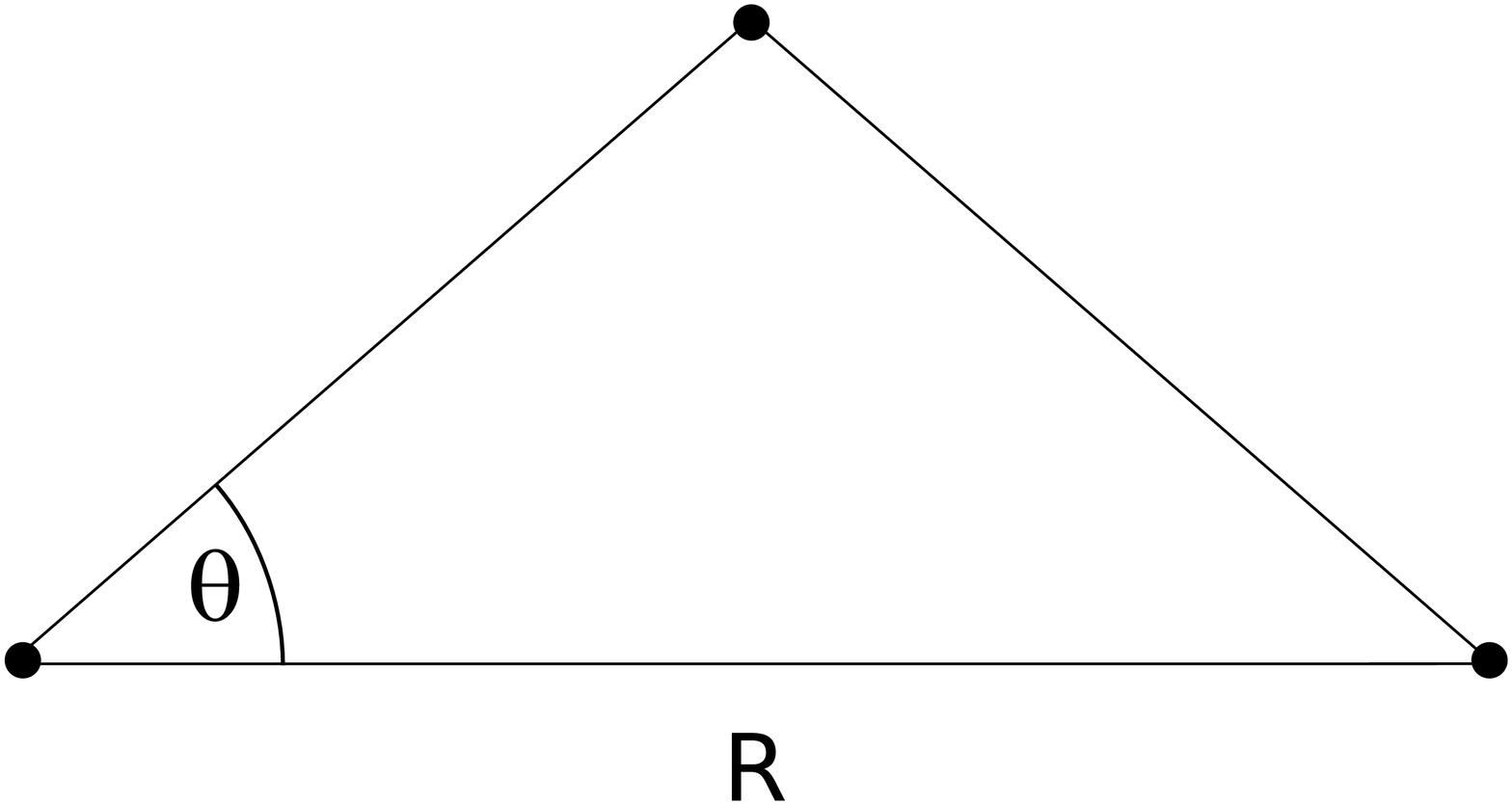}
  \caption{Parameters specifying the isosceles triangle configurations of \ce{Ne3} and~\ce{Ar3}.}
  \label{isosceles-triangle}
\end{figure}

\begin{figure*}[tb]
  \includegraphics[width=\textwidth]{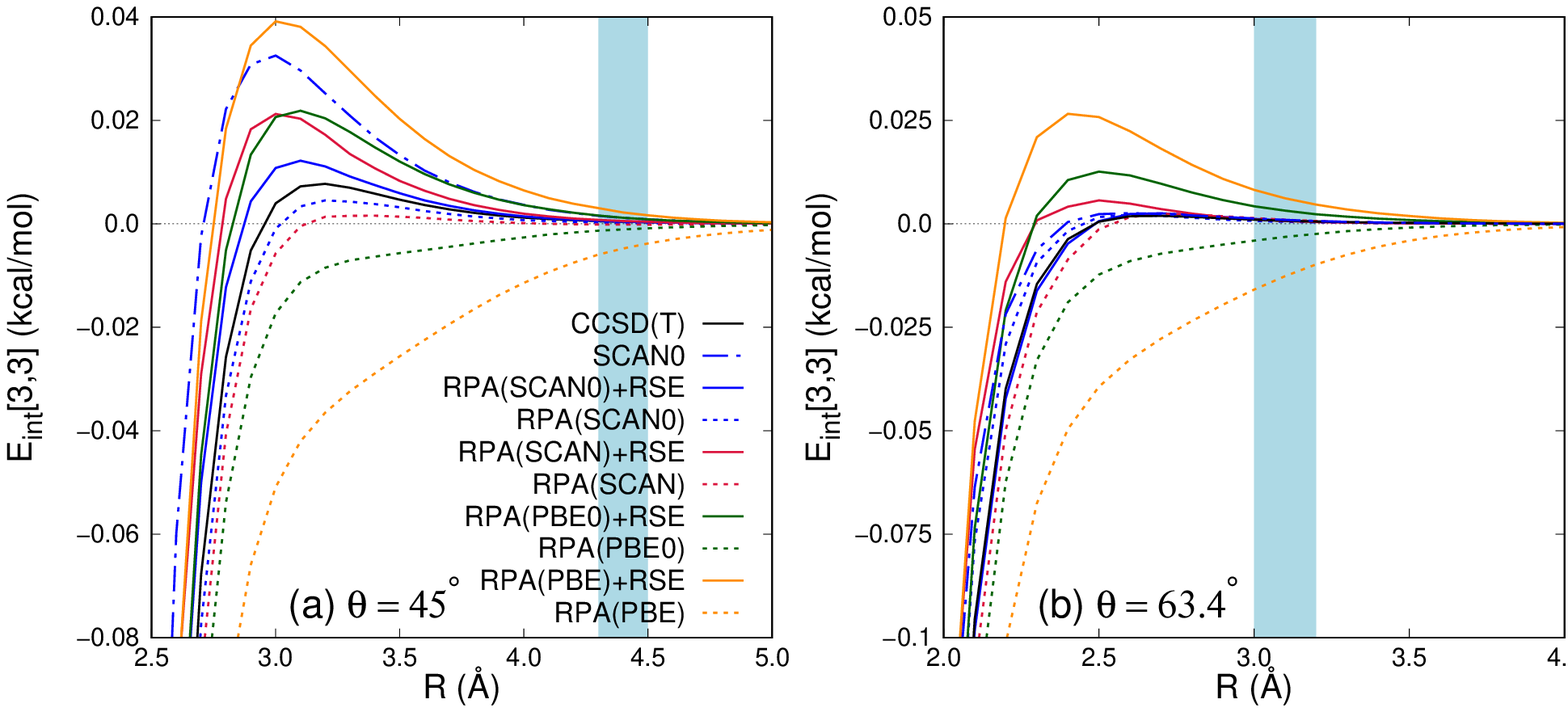}
  \caption{Nonadditive three-body interaction energy of neon trimers.
    The blue background denotes configurations where the closest pair of atoms is within $\pm0.1~\AA$
    of the equilibrium dimer separation. 
  } \label{fig:Ne3}
\end{figure*}

\begin{figure*}[tb]
  \includegraphics[width=\textwidth]{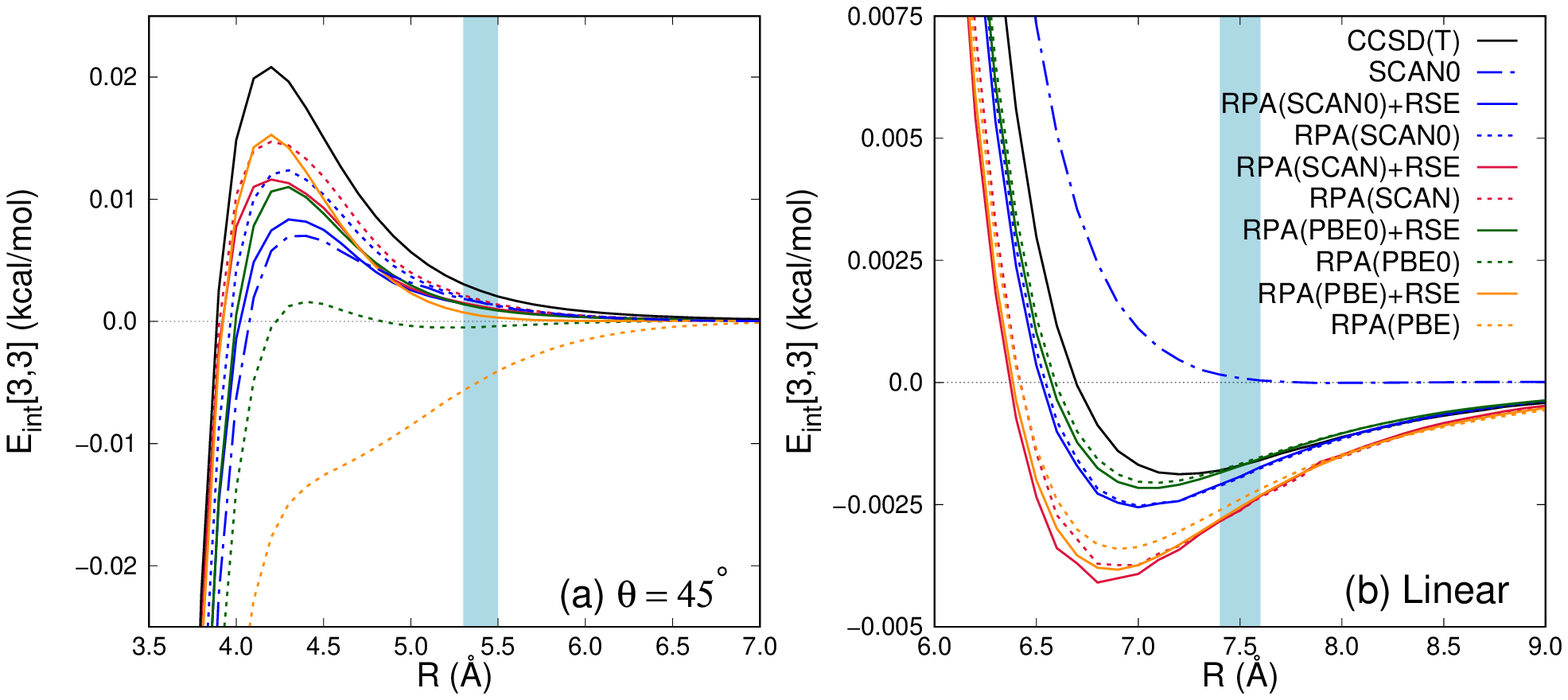}
  \caption{Nonadditive three-body interaction energy of argon trimers.
      The blue background denotes configurations where the closest pair of atoms is within $\pm0.1~\AA$
    of the equilibrium dimer separation. } \label{fig:Ar3}
\end{figure*}

\begin{figure*}[tb]
  \includegraphics[width=0.9\textwidth]{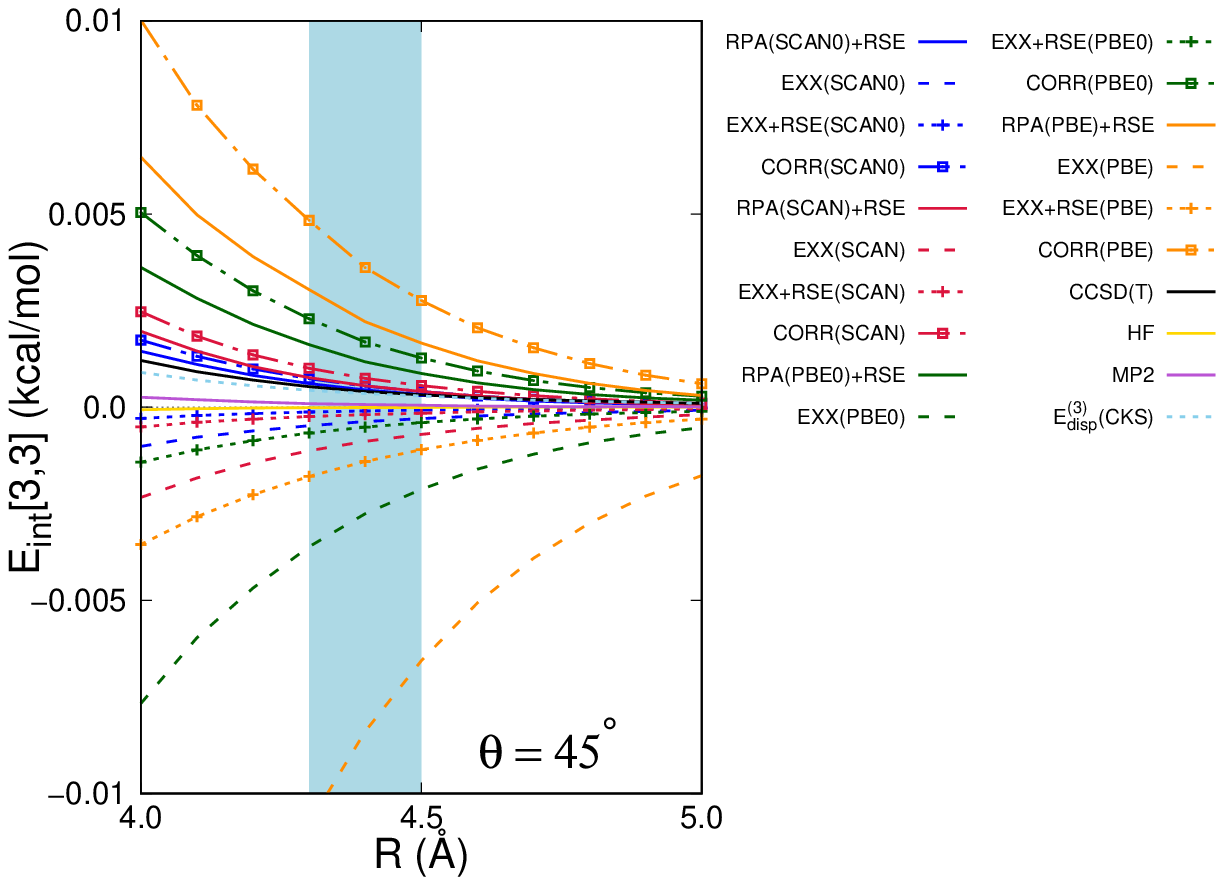}
  \caption{EXX component of the nonadditive RPA interaction energy for \ce{Ne3} at \ang{45}. $R$ is the length
    of the isosceles triangle base. The blue background denotes configurations where the closest pair of atoms is within $\pm0.1~\AA$
    of the equilibrium dimer separation.} \label{fig:Ne3EXX}
\end{figure*}

The statistical measures of the RPA results
indicate that the most challenging trimers for RPA(DFT) and RPA(DFT)+RSE
are weakly interacting systems with a high share of the dispersion energy.
To better understand the source of RPA's errors for those systems,
as well as the sensitivity to the input orbital set,
we turn to a case study of noble gas trimers.

The systems we consider are trimers of neon and argon
in different configurations.
Specifically, we use two isosceles triangle configurations with angles
of 45 and 63.4 degrees and a linear configuration.
See Figure~\ref{isosceles-triangle}
for the definition of the geometric parameters. 
The linear configurations correspond to
a negative asymptote of the three-body dispersion energy; the remaining configurations
correspond to a positive  asymptote.
The geometries and tabulated numerical data are included in the Supporting Information.

The nonadditive interaction energy curves are shown in Figures~\ref{fig:Ne3} and~\ref{fig:Ar3}
for neon and argon, respectively.
One can see that the performance of RPA(DFT) clearly depends on the input orbitals.
The RPA(PBE) scheme, with and without RSE, clearly performs the worst
with the interaction energy curve visibly further from
the reference compared to the remaining methods.
RPA based on PBE0 is of poorer quality than the approaches
with the SCAN and SCAN0 orbitals, but it achieves a similar
accuracy once the RSE correction is included.
For RPA(PBE) and RPA(PBE0), the RSE correction is necessary
to reproduce the local maxima on the energy curves.
In contrast, RPA(SCAN) and RPA(SCAN0) are qualitatively
correct even without RSE.
Quantitatively, RPA(PBE0)+RSE and the approaches based
on the SCAN and SCAN0 achieve a similar level of accuracy.

We show the data obtained with SCAN0 for comparison in Figures~\ref{fig:Ne3} and~\ref{fig:Ar3}.
It performs rather well for the triangle configurations where there is a 
density overlap of the three atoms.
However, it lacks the attractive three-body dispersion interaction needed to describe the 
binding curve of the trimer.

  \begin{table}
    \caption{Ionization potentials (eV) for isolated noble gas atoms, computed with Koopmans's theorem.
      The experimental IPs are taken from the NIST Atomic
      Spectra Database.\cite{NIST_ASD}} \label{koopmans}
    \begin{tabular}{lrr}
        \toprule
        Method & \ce{Ne} & \ce{Ar} \\
        \midrule        
        SCAN0 & 16.5 & 12.3 \\
        PBE0 & 16.0 & 12.0 \\
        SCAN & 14.0 & 10.7 \\
        PBE & 13.3 & 10.3 \\
        exp. & 21.6 & 15.8 \\ 
        \bottomrule
    \end{tabular}
  \end{table}

To rationalize the differences between various orbital sets, we note that
the long distance decay of the electron density is controlled by the ionization potential (IP),
which equals, by Janak's theorem, negative the HOMO eigenvalue.
As seen in Table~\ref{koopmans}, the IPs at the DFT level approach the experimental
values from below.
Otherwise stated, all DFT methods yield electron densities which are
too diffuse, and significantly so because the errors in the IPs
are on the order of tens of percent.

The ordering of methods in terms of increasing
IPs is $\text{PBE} < \text{SCAN} < \text{PBE0} < \text{SCAN0}$.
That sequence correlates with the magnitude of
the RPA nonadditive interaction energy components (Figure~\ref{fig:Ne3EXX}).
Specifically, the magnitudes of the EXX energy, singles correction,
and RPA correlation energy form the sequence $\text{PBE} < \text{PBE0} < \text{SCAN} < \text{SCAN0}$.
Interestingly, the SCAN and PBE0 are reversed with respected to the order given 
by the IPs.
We have computed the atomic density and found that, for the distance of interest, 
the atomic densities of PBE0 and SCAN are almost identical.
Thus, the addition of Hartree-Fock exchange reduces the errors for both PBE an SCAN
and using SCAN instead of PBE reduces the errors related to too delocalized states.

We now argue that, for the considered systems, the decay rate
of the RPA energy components is an indication of the quality
of the orbitals provided to the RPA energy formula.  
Because the EXX part of the RPA energy is based on the orbitals
obtained with a hybrid or a semilocal DFT model, we expect it
to account only for the physical terms that depend on the density overlap.
Here, that would be the exchange nonadditivity and the
intramonomer correlation corrections to it.\cite{chalasinski1990calculations,chalasinski1994supermolecular}
By the above reasoning, the decay rate of the EXX components should be akin
to that of HF and MP2, which respectively describe the abovementioned terms.
However, we observe that the magnitude of the EXX energy is, for all tested orbital sets,
significantly larger than that of HF and MP2 at long range and has a sign
opposite to the MP2 energy (Figure~\ref{fig:Ne3EXX}).
Therefore, the large magnitude of EXX seen here is an artifact of approximate DFT
functionals.
 
The excessive EXX term is partially cancelled by the RPA correlation,
which has to be much larger in magnitude than the accurate
three-body dispersion for the compensation to occur.
The singles correction partakes in the cancellation of EXX,
but does not remove the artifact entirely.
A reliance on the cancellation of unphysical contributions
appears to deteriorate the results for RPA(PBE) and RPA(PBE)+RSE,
as those are the methods with the largest amount of cancellation
between the different terms and also the worst performers for the noble gas trimers.

\begin{table}[bt]
  \caption{Effect of the RSE and GW singles correction on the nonadditive
    interaction energy (kcal/mol) of \ce{Ne3} at angle $\ang{45}$ and $R=3.0~\AA$.} \label{tab:gw}
  \begin{tabular}{lr}
    \toprule
    Method & $E_\text{int}[3,3]$ \\
    \midrule
    CCSD(T)        &  0.0040 \\
    RPA(SCAN)      & $-$0.0057 \\
    RPA(SCAN)+RSE  &  0.0213 \\
    RPA(SCAN)+GWSE &  0.0163 \\
    RPA(SCAN0)     & $-$0.0009 \\
    RPA(SCAN0)+RSE &  0.0108 \\
    \bottomrule
  \end{tabular}
\end{table}

A step beyond RPA(DFT)+RSE would be an application of the GW singles correction described
in Ref.~\citenum{klimes2015singles} and applied, e.g., for the phase diagram of ice\cite{zen2018fast}
and for the binding energy curve of water on graphene.\cite{brandenburg2019physisorption}
While due to technical reasons it is currently not possible
to run large scale computations of GWSE for molecules using VASP,
we have computed GWSE for a single neon trimer at $\theta=\ang{45}$ and $R=3.0~\AA$
to probe its effect for the systems considered in this work (see Table~\ref{tab:gw}).
For RPA(SCAN), the difference between GWSE and RSE corrections is
on the same order of magnitude as the reference interaction energy
at the considered distance and comparable to the effect of changing
the orbital set from SCAN to SCAN0.
A futher investigation of the GWSE correction remains a subject of our future work.

\section{Conclusions}

We have examined numerical and theoretical aspects of applying
RPA for many-body noncovalent systems of atoms and molecules. 
We introduced a cubic scaling algorithm for molecular RPA which achieves
high and controllable numerical precision.
Unlike prior efficient RPA implementations, the cubic scaling does not
assume the sparsity of the effective density matrices, or Green's functions.
It employs a systematically improvable Cholesky basis instead of
the usual auxiliary basis sets for the decomposition of the Coulomb matrix.
Those features make the algorithm fit for
accumulating subtle $n$-body contributions in clusters
of interacting molecules.

Regarding the accuracy of nonadditive interaction energies at the RPA level,
the choice of orbitals affects RPA quantitatively and, in some cases, qualitatively.
To assess RPA's dependence on the Kohn-Sham state, we tested
GGA and meta-GGA exchange-correlation models: PBE, PBE0, SCAN, and SCAN0.
In addition, we tested the singles correction, RSE,
which effectively changes the electron density, affects the Hartree-Fock part
of the RPA energy, but does not affect
the RPA correlation contribution.

Our statistical data on the 3B-69 set of trimers demonstrate that the best RPA variants
are based on SCAN0 hybrid meta-GGA (applied without RSE)
and PBE0 hybrid GGA (applied in combination with RSE).
The RPA methods achieve a much better accuracy than their base DFT functionals.
Compared to wavefunction methods, the accuracy of RPA(SCAN0) and RPA(PBE0)+RSE is
between the MP3 and CCSD  approaches, 
which have orders of magnitude larger requirements of storage and compute time.
For solid state calculations hybrid functionals incur additional computational cost.
Our data for noble gas trimers suggest that when hybrid DFT calculations
are not feasible, SCAN is currently the best choice of a pure DFT model.
Following the standard practice, that is using RPA with PBE orbitals,
gives the worst predictions over the entire 3B-69 dataset.
The results for noble gas trimers suggest that the reason behind the poor performance of
RPA(PBE) and RPA(PBE)+RSE is that the Hartree-Fock part of the RPA interaction energy
in those cases decays at an artificially slow rate as a function of the intermonomer distance.

Out of two major advantages that RPA has over semilocal DFT, that is, the account of the dispersion energy
and the compatibility with exact exchange, the latter appears to be especially important as
it eliminates the artificial exchange overlap interactions already reported in the DFT literature.\cite{gillan2014many}
As a result we observe a near-benchmark accuracy of our best RPA variants for the low-dispersion subset 
of the 3B-69 dataset.

Overall, we find that RPA is, in terms of computational cost and accuracy, a well balanced scheme
for predicting many-body energies of systems bound by noncovalent interactions.
Its accuracy to cost ratio makes it a method preferable to both hybrid semilocal DFT
and simple wave function approaches, e.g., MP2 and MP3.
We have identified that the DFT errors of the base functional visibly transfer to the RPA results,
which implies that a further improvement of the RPA methodology is still possible by
devising better schemes for generating the orbital input.

\begin{acknowledgement}
This work was supported by the European Research Council (ERC)
under the European Union's Horizon 2020 research and innovation 
program (grant agreement No 759721).
We are grateful for the computational resources provided by the 
IT4Innovations National Supercomputing Center (LM2015070),
CESNET (LM2015042), and CERIT-SC (LM2015085) funded within the programme 
``Projects of Large Research, Development, and Innovations Infrastructures''
of the Ministry of Education, Youth, and Sports.
This research was supported in part by PLGrid Infrastructure.
We thank David P. Tew for help with obtaining initial results.
We thank Aleksandra Tucholska for giving us access to her coupled-cluster program
for generating electronic densities.
\end{acknowledgement}

\begin{suppinfo}
Spreadsheets with raw numerical data and computational details,
geometries in a form of xyz files. The supporting information
is available free of charge via the Internet at https://pubs.acs.org/doi/10.1021/acs.jctc.9b00979 .
\end{suppinfo}

{\footnotesize
\bibliography{biblio}}

\end{document}